\documentclass[prd,10pt,twocolumn]{revtex4} 
\usepackage{graphicx}   
\usepackage{hyperref}   

\begin{document}

\title{Excited-state decay in strictly Everett-like interpretations of quantum mechanics} 
\author{Jon Geist}
\date{2012 September 15}
\email{jon.geist@ieee.org.}

\maketitle

\section*{abstract}
Excited state decay is examined within the framework of strictly Everett-like formulations of quantum mechanics.  Even though these formulations were developed to treat particles and systems of particles as part of a larger system that includes a measurement apparatus, the analysis is carried out in terms of isolated particles because excited state decay measurements are performed under conditions that approximate isolation.  It is shown that the time evolution of the wave function describing each particle in a sample of well-isolated identical particles in their lowest excited state must satisfy the same branch topology for all strictly Everett-like interpretations.  This topology describes a countably infinite sequence of random branching events that occur at a rate $\lambda_B$ for each particle in the sample.  Two more parameters are required to describe excited-state decay measurements.  The first, $\lambda_A$, is the expectation value of the excited-state decay rate determined by an observer state measuring the lifetime of a sample of isolated identical particles.  The second, $\varepsilon$, is the probability that an observer state that branches off an observer state associated with the excited state is also associated with the excited state following the branching event.  In strictly Everett-like formulations, $\varepsilon$ must be independent of time and branch, but there does not appear to be any way to determine its value within the framework of the formulations.  Therefore, contrary to Everett's goal, strictly Everett-like interpretations do not provide a complete description of every system.  However, it is possible to show that $\lambda_A = (1-\varepsilon) \lambda_B$, which with the agreement between theoretical and experimental lifetime determinations provides a conservative upper limit of $\varepsilon \alt$ 0.1\% at the current state of the art. This result is very different from the example of a superposition of states originally considered by Everett wherein the probability of each final state is given by the conventional Born's rule and no new parameters are required.  

\section{\label{Intro}Introduction} 

In 1957 Everett proposed a formulation of quantum theory in which \lq \lq a wave function that obeys a linear wave equation everywhere and at all times supplies a complete mathematical model for every isolated physical system without exception ... and every system that is subject to external observation can be regarded as part of a larger isolated system" that includes the observer \cite{Everett}. His aim was \lq \lq not to deny or contradict the conventional formulation of quantum theory, ..., but rather to supply a new, more general and complete formulation, from which the conventional interpretation can be deduced." \cite{Everett} 

Everett's formulation introduced a radical new interpretation of quantum theory: \lq \lq Thus with each succeeding observation (or interaction), the observer state \lq  branches' into a number of different states" where \lq \lq all elements of a superposition (all branches) are \lq actual', none any more real than the rest." \cite{Everett} Furthermore, he concluded that the results of his \lq \lq formalism agree with those of the conventional \lq external observation' formalism in all those cases where that familiar machinery is applicable." \cite{Everett}  However, he did not explicitly consider the application of his formulation to excited state decay, confining his treatment to superpositions of states.  

Because Everett never clearly explained how to implement his formulation, his work has motivated several different elaborations \cite{Schlosshauer}.  Nevertheless, all Everett-like interpretations share two distinguishing characteristics.  First, \lq \lq every term in the final-state superposition of quantum states represents an equally  \lq real' physical state of affairs that is realized in a different branch of reality'' \cite{Schlosshauer}. This characteristic distinguishes Everett-like interpretations from the conventional interpretation of quantum theory, which predicts the relative probabilities of alternative realities but posits truly random processes that actualize a single reality among the multiple possibilities.  Second, a single wave equation that supplies a complete mathematical description of every isolated physical system distinguishes Everett-like interpretations from local hidden variable \cite{Mermin} theories.  

The purpose of this report is to study the Everett-like description of the transition (decay) of a particle from its lowest excited state to its ground state in terms of the network topology of branching realities and to determine the constraints imposed by this topology. (As used here, network topology is the study of the connections among the branches of a network without regard to their length.)  For simplicity, the network topology will be referred to as the topology, an isolated particle in its lowest excited state will be referred to as a particle, and the transition of an isolated particle from its lowest excited state to its ground state will be referred to as a decay event unless clarity requires a more complete description.   

The analysis will be carried out in terms of a sample containing a large number of isolated particles and will determine the single-particle transition rate $0 < \lambda_A < \infty$ for decay events as determined by any observer state in terms of the single-particle transition rate $0 < \lambda_B < \infty$ at which reality branching occurs in a strictly Everett-like description of an isolated particle.  This approach is justified because it is possible to conduct excited-state decay measurements on samples that include observer states but behave very much as if the particles making up the sample were isolated \cite{Volz2} \cite{Volz} as described in more detail in the discussion section.  

\section{\label{Branching}Excited State Decay} 

According to the conventional formulation of quantum theory, the probability that a single isolated particle initially in its lowest excited state at time $t=0$ will still be in its lowest excited state at time $t >0$ is given by the survival function
\begin{equation}
S(\lambda t) = e^{-\lambda t}
\label{S(t)} 
\end{equation}
where $\lambda > 0$ is the transition rate for the decay process, which is well approximated by 
\begin{equation}
\lambda = \frac{2 \pi}{\hbar} \vert V_{eg}(E) \vert^2 \rho(E),
\label{lambda}
\end{equation}
when $\vert V_{eg}(E) \vert^2 \rho(E)$ is a slowly varying function of $E=E_e-E_g$, where $E_e$ and $E_g$ are the respective excited state and ground state energies of the particle, $V_{eg}(E)$ is the matrix element of the time independent perturbation inducing the transition, and $\rho(E)$ is the energy density of the ground state and associated decay products \cite{Fischbach}.  For the rare cases where $\vert V_{eg}(E) \vert^2 \rho(E)$ is not a slowly varying function of $E$, more accurate expressions than Eq. \ref{lambda} can be derived \cite{Fischbach}.  With this caveat Eqs. \ref{S(t)} and \ref{lambda} are supported by an overwhelming body of theoretical and experimental results.  Therefore, any Everett-like description of excited state decay must be consistent with these equations.  

Because the isolated particle is in its lowest excited state, only two branches can be created by a decay event, one in which the particle remains in its excited state and one in which the particle decays to its ground state. Furthermore, $\lambda_A$ must be independent of time and well approximated by Eq. \ref{lambda} to satisfy Everett's goal to avoid denying or contradicting the conventional formulation of quantum theory. Finally, $\lambda_B$ must be proportional to $\lambda_A$ to avoid adding a new time-dependent equation to the conventional theory.  Thus, $\lambda_B$ is also independent of time in what follows.  

\begin{figure}[htbp] 
  \centering
 \includegraphics[width=3.8in,height=3.0in,keepaspectratio]{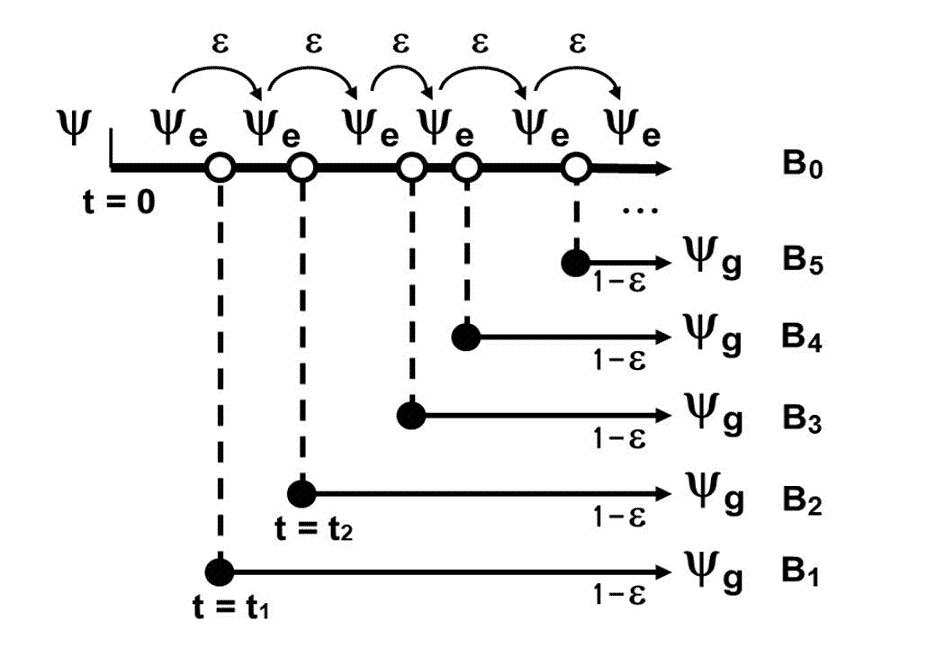}
  \caption{The only possible branching topology in a strictly Everett-like interpretation for a wave function $\psi$ that describes the time evolution of an isolated particle in its lowest excited state at time $t=0$, where $\psi_g$ and $\psi_e$ are the wave functions that describe the ground state and the lowest excited state of the particle, respectively.  The open circles indicate that a branching event that cannot be observed by an observer state on the branch \textbf{B$_0$} occurred on that branch.  The filled circles indicate that a branching event that can be observed as a decay event occurred on branch \textbf{B}$_{i>0}$.  The quantities $\varepsilon$ are the excited state branching probabilities defined in the text.}  
\label{topology}
\end{figure}
 
These constraints are sufficient to eliminate all candidate topologies for the time evolution of an Everett-like wave function that describes the state of a particle in its lowest excited state except for that illustrated in Fig. \ref{topology}.  In this figure $\psi$ describes the time evolution of a single particle that is described at time $t_0$ by $\psi_e$, where $\psi_e$ describes the first excited state of the particle.  The different branches are labeled \textbf{B$_i$} in the figure, where $i-1$ decay events precede the decay event that creates branch \textbf{B}$_i$ ($i > 0$).  

A filled circle in Fig. \ref{topology} indicates that a decay event occurred on the branch following the filled circle.  An open circle indicates that a branching event occurred during the time evolution of the wave function on the branch following that open circle, but that a decay event did not accompany the branching process on that branch.  A thick horizontal line indicates that the transition rate for the branching process is a positive constant given by Eq. \ref{lambda} and a thin horizontal line indicates that the transition rate is zero. 

The quantity $\varepsilon$ in Fig. \ref{topology} is the excited-state branch probability, which is defined as the probability that an observer state that branches off an observer state associated with the excited state is associated with the excited state following the branching event.  To satisfy Everett's requirement that a wave function that obeys a linear wave equation everywhere and at all times supplies a \textit{complete} mathematical model for every isolated physical system, $\varepsilon$ must be independent of the branch index $i$ as assumed in the figure.  

At time $t=t_1$ in Fig. \ref{topology} a decay event occurs and the wave function $\psi$ branches into two branches.  The branch \textbf{B}$_0$ is still described by the wave function $\psi_e$ and the transition rate on this branch is still given by $\lambda_B$.  The other branch \textbf{B}$_1$ is described by $\psi_g$, where $\psi_g$ describes the ground state of the particle.  The transition rate on this branch is zero because the particle is isolated so there is no source of energy to excite it to a higher energy state.  At a later time $t_2$, another decay event occurs on branch \textbf{B}$_0$.  The branch \textbf{B}$_0$ is still described by the wave function $\psi_e$, the transition rate on that branch is still given by $\lambda_B$, and the other branch, \textbf{B}$_2$, is described by $\psi_g$ with zero transition rate.  This process continues indefinitely because $\lambda_B > 0$ for all times $t>0$.       

The details of how the wavefunction $\psi$ is related to the particle wave functions $\psi_e$ and $\psi_g$, the wave functions of the other decay products of the decay process (photons for electronic states, etc.), the observer state wave functions, as well as the detailed nature of the branches all depend upon the particular strictly Everett-like being considered \cite{Barrett}\cite{Vaidman}\cite{Schlosshauer}.  These details are beyond the scope of this work. However, the topology of the branching of the wave functions for all strictly Everett-like formulations must conform to that shown in Fig. \ref{topology} to be consistent with the conventional formulation of quantum theory as required by Everett.

The branching process illustrated in Fig. \ref{topology} is supposed to be continuous even though the figure may create the false impression that it is a discontinuous process. When the entire topology of Fig. \ref{topology} is viewed, which has been called the outside view \cite{Tegmark1}, the wave function $\psi$ appears to be growing new branches while evolving continuously.  On the other hand, from the inside view \cite{Tegmark1} of an observer state associated with any branch \textbf{B}$_{i>0}$, a single discontinuous change from the excited state described by $\psi_e$ to the ground state described by $\psi_g$ occurs at some time.    

At the current state of the art of experimental physics, an observer state associated with a branch \textbf{B}$_{i>0}$,  which is able to observe the single particle decay event on that branch, is unable to observe the single particle decay events on any other branches or even how many other branches simultaneously exist at any given time. Furthermore, an observer state that is associated with the wave function $\psi_e$ on branch \textbf{B}$_0$ is incapable of observing any decay event at all.  There are two important consequences of these constraints.  First, an observer can detect at most one decay event per particle in accordance with current experimental facts.  Second, if $\varepsilon > 0$, the average rate $\lambda_B$ at which branching events occur will be greater than the expectation value of the rate $\lambda_A$ at which excited state decay events are perceived to occur by any given observer state observing a sample containing a large number of particles.  

The fact that strictly Everett-like formulations of excited state decay requires two different transition rates, whereas the conventional formulation requires only one, as well as the existence of the the branch probability $\varepsilon$, is a major difference between the two formulations.  If the former is to be a complete theory, it must be possible to derive the dependence of $\lambda_A$ and $\varepsilon$ on $\lambda_B$ within the framework of the formulation.  

\section{\label{Relation}The relation between $\lambda_A$, $\varepsilon$, and $\lambda_B$}

To proceed, let 
\begin{equation}
W = 1/\lambda_B
\label{B}
\end{equation}
be the expectation value of the waiting time \cite{Pitman} between adjacent branching events in the topology of Fig. \ref{topology} and let $W_i$ be the expectation value of the waiting time for a decay event to occur on any branch B$_{i>0}$.  Because $\lambda_B$ must be independent of time to be consistent with the conventional formulation of quantum theory, 
\begin{equation} 
W_i = i W, 
\label{T_i}
\end{equation}
while the actual waiting times $t$ from time $t=0$ for a decay event to occur on branch \textbf{B}$_{i>0}$ is randomly drawn from the probability density function \cite{London} 
\begin{equation}  
f_i(\lambda_B, t) = \lambda_B e^{-\lambda_B t} \frac{(\lambda_B t)^{i-1}}{(i-1)!},          
\label{f_i}         
\end{equation}
which is the time derivative of the cumulative distribution function \cite{London}
\begin{equation}
F_i(\lambda_B t) = 1 - S_i(\lambda_B t),
\label{F_i} 
\end{equation}
where 
\begin{equation} 
S_i(\lambda_B t) = e^{-\lambda_B t} \sum_{n=1}^i \frac{(\lambda_B t)^{n-1}}{(n-1)!}     
\label{S_i}
\end{equation}
is the survival function \cite{London} of the excited state on branch \textbf{B}$_i$. \cite{Pitman} 

\begin{figure}[htbp] 
  \centering
 \includegraphics[width=4in,height=3in,keepaspectratio]{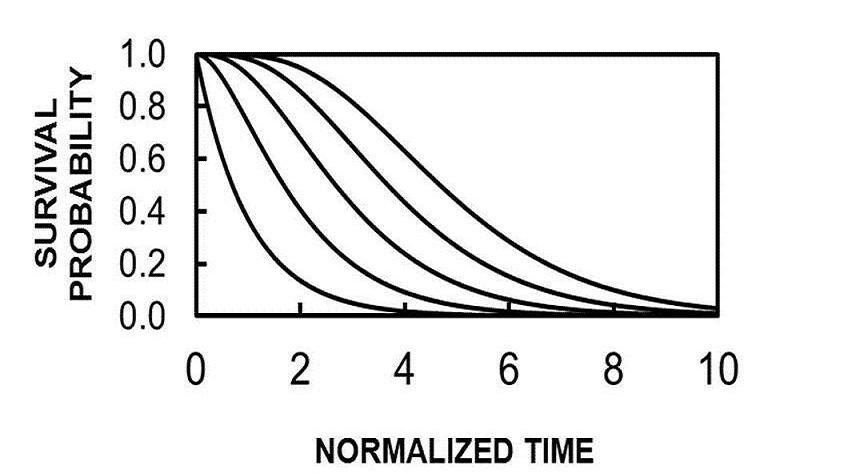}
  \caption{The survival function $S_i(\lambda_B t)$ of the excited state on the different branches \textbf{B}$_{i>0}$ in Fig. \ref{topology} as a function of normalized time $t/W$, where  $W$ is the average waiting time between branching events in the outside view of the branching topology of a single isolated particle decaying from its first excited state to its ground state and $\lambda_B = 1/W$.  From the inside view \cite{Tegmark1} only branch B$_1$ decays exponentially.}  
\label{survival}
\end{figure}

Figure \ref{survival} plots the survival function $S_i(\lambda_B t)$ of the excited state on branch $i$ for $1 \le i \le 5$ from the normalized time $t/W=0$ to $t=6$.  Only the survival function on branch \textbf{B}$_1$ is exponential. Furthermore, it is only from the inside view \cite{Tegmark1} that the quantities $S_i(\lambda_B t)$ are survival functions because it is only from this view that the excited state disappears.  From the outside view, the excited state survives indefinitely, so from this view, $S_i(\lambda_B t)$ describes the probability that the decay event resulting in branch $i$ will not occur before time $t$ rather than a true survival function.  But the mathematics is identical for both views.    

Consider a sample of $N$ isolated identical particles in their lowest excited states.  The time evolution of the wave function that describes each of these particles must follow the topology given in Fig. \ref{topology}.  Let C$_i$ denote the set that contains all of the branches with survival functions given by $S_i(\lambda_B, t)$.  If an observer were able to choose the same branch on which to observe the decay event of each particle in the sample, then this observer would be able to measure each of the survival functions plotted in Fig. \ref{survival}.  However, this is not possible at the current state of the art.  Instead, the branch on which an observer measures a decay event for each particle in the sample appears (at the current state of the art of experimental physics) to be selected randomly. 

To determine the expectation values of the number of branches of each class that contribute to the apparent transition rate $\lambda_A$ measured by any given observer, verify that the expectation value for the number of observer state branches in set C$_i$, is given by 
\begin{equation}
N_i = N (1-\varepsilon) \varepsilon^{i-1}.
\label{N_i}  
\end{equation}

If $\varepsilon = 0$ then $N_i = 0$ for $i > 1$ and there are no observer states on branch \textbf{B}$_0$ following the first decay event, so this case is no different from conventional quantum theory.  On the other hand, if $\varepsilon = 1$, then it is still the case that $N_i = 0$, but there are no observer states to detect decay events on any branches, so no decay events are ever observed, which contradicts experimental fact.  Each case violates one of Everett's goals, so $0 < \varepsilon < 1$ in what follows.  

For large $N$, lifetime measurements will give a good approximation to the weighted average expectation value of the apparent lifetime 
\begin{equation} 
\tau_A = \frac{1}{N} \sum_{i=1}^\infty N_i W_i = \frac{W}{1-\varepsilon}.  
\label{Tau_A} 
\end{equation}
(The derivation of the right most expression in this equation and the derivations of the right most expressions in Eqs. \ref{D_A} and \ref{P_A} are given in the Appendix.)  

Similarly, from Eq. \ref{f_i} the distribution from which the apparent lifetimes are drawn is given in terms of the branching rate $\lambda_B$ by the weighted average distribution
\begin{equation}  
f_A(\lambda_B, t) =  \frac{1}{N}\sum_{i=1}^\infty N_i f_i(\lambda_B, t) 
			= f_1(\lambda_A,t),  
\label{D_A}         
\end{equation}
where $\lambda_A$ is defined as 
\begin{equation}
\lambda_A = (1-\varepsilon) \lambda_B
\label{lambda_A}
\end{equation}

Finally, from Eq. \ref{S_i} the apparent survival function of the particles is given by  
\begin{equation} 
S_A(\lambda_B t)  = \frac{1}{N} \sum_{i=1}^\infty N_i S_i(\lambda_B t) = e^{-\lambda_A t} = S_1(\lambda_A t).
\label{P_A}     
\end{equation} 
 
According to Eqs. \ref{Tau_A}, \ref{B}, and \ref{lambda_A}, 
\begin{equation}
\tau_A = \frac{1}{\lambda_A},
\label{Tau_AX}       
\end{equation}
which is a necessary and sufficient condition for Eq. \ref{P_A} to have the mathematical properties of a survival function, where $\lambda_A$ and $\tau_A$ are the experimentally determined transition rate and lifetime of the excited state, respectively.  

The validity of Eqs. \ref{Tau_A}-\ref{Tau_AX} is independent of the value of the excited-state branch probabilities 
$\varepsilon$, so there is no relation between the transition probability calculated from Eq. \ref{lambda} and $\varepsilon$.  This is very different from the example considered by Everett, in which the branch probability of the $j^{th}$ branch is given by the conventional Born's rule,
\begin{equation}
\epsilon_j = |(\psi_i,\phi_j)|^2,
\label{epsilon_j}
\end{equation}
where the initial state is the superposition of eigenstates $\phi_j$
\begin{equation}
\psi_i = \Sigma \epsilon_j \phi_j.  
\end{equation}

In the conventional formulation of quantum theory, the matrix element in Eq. \ref{lambda} describes a physical process with an exponential survival function.  On the other hand, in strict Everett-like formulations studied here, there is no physical process with an exponential survival function that has a transition rate of $\lambda_A$.  Instead, $\lambda_A$ is a weighted-average transition rate computed over an infinite number of physical decay events occurring on different branches of the topology given in Fig. \ref{topology}. 
 
Only the single particle decay event that gives rise to branch \textbf{B}$_1$ has an exponential survival function, and its transitions rate is $\lambda_B = \lambda_A/(1 - \varepsilon)$ rather than $\lambda_A$.  For Eq. \ref{lambda} to describe $\lambda_A$, a new derivation of that equation involving new physics is necessary, but this would violate Everett's goal \lq \lq not to deny or contradict the conventional formulation of quantum theory".  

On the other hand, if Eq. \ref{lambda} describes $\lambda_B = 1/W$ rather than $\lambda_A$, which is completely consistent with the conventional formulation of quantum theory, the excited state branch probability $\varepsilon$ must be small enough that the experimentally determined value of $\lambda_A$ cannot be distinguished at the current state of the art from the theoretically determined value of $\lambda_B$.   

\section{\label{Discussion}Discussion} 

The above analysis was carried out in terms of a sample containing a large number of isolated particles, even though observation of the decay events requires that observer states be included in the full description of the system, which by definition precludes isolation of the particles.  However, this approach is justified by the fact that it is possible to conduct excited state decay measurements in a wide variety of systems that include observer states but behave very much as if isolated with respect to the decay process being observed \cite{Volz2} \cite{Volz} before the decay-product states are coupled by decoherence \cite{Zurek} with states that describe portions of the measurement apparatus.  Furthermore, it is possible to vary key experimental parameters to allow extrapolation of the lifetime or transition probability to that of an isolated particle. \cite{Volz2}   

In conclusion, within the strict framework of Everett's original program, there is only one branching topology and three pertinent quantities $\lambda_B$, $\varepsilon$, and $\lambda_A$ that describe the decay to the ground state of an isolated particle initially in its lowest excited state.  Only two of these quantities are independent.  The branching rate $\lambda_B$ is determined by Eq. \ref{lambda}, $\lambda_A$ is determined experimentally, and $\varepsilon$ can then be determined from Eq. \ref{lambda_A}, which currently provides an upper limit for $\varepsilon$ of less than or approximately equal to 0.1\%.  However, the striking asymmetry between the branch probabilities $\varepsilon$ and $1-\varepsilon$ highlights \lq \lq the well-known problem of the perplexing feature that, loosely speaking, some observers are more equal than others" recently addressed in \cite{Tegmark2}.  

However, an intuitively satisfying solution to this problem has already been proposed for the case of a superpositions of states.  According to this addition to Everett's formulation each branch is associated, not with a single observer state, but with a number of observer states where the number on any branch is proportional to the Born's rule probability $|(\psi_i,\phi_j)|^2$. \cite{Deutsch} This assumption assures the \lq \lq equality" of all observers of a \lq \lq collapse of the wavefunction" from a superposition of states to a single eigenstate.  But it is not intuitively satisfying for excited state decay because, unlike the case of superposition, the value of branch probabilities are not related to any quantities in conventional quantum theory and must be determined experimentally.  
 
Consequently, and contrary to his stated goal, Everett's formulation is not a complete formulation of quantum theory.  Furthermore, descriptions of excited state decay that are completely consistent with strictly Everett-like formulations appear to challenge his goal that no branches are any more real than the rest.  On the other hand, if future work were to accurately determine a non-zero value of $\varepsilon$, this would constitute experimental confirmation of a prediction of strictly Everett-like formulations.  Such a result would be very significant because it would appear to rule out the conventional interpretation of quantum theory. On the other hand, failure to detect a non-zero Everett branch probability will not rule out strictly Everett-like formulations, just as failure to measure a non-zero neutrino mass does not rule out the possibility of a neutrino rest mass.  However, it would probably reduce the appeal of strictly Everett-like interpretations as a replacement for the conventional formulation and interpretation of quantum theory. In either case, it appears likely that excited state decay will provide better tests of extensions of Everett-like formulations than superpositions of states.

\begin{acknowledgments}
The author thanks Samuel M. Stavis, Stefan Leigh, and Richard A. Allen for helpful discussions, suggestions, and comments.  
\end{acknowledgments}

\appendix

\begin{widetext} 

\section{\label{Unending}Derivation of equations} 

The purpose of this appendix is to describe the derivation of eqs. \ref{Tau_A}-\ref{P_A} in the body of this report. To start, assume that $0 \le \varepsilon \le 1$, let
\begin{equation}
\beta= \frac{\varepsilon}{1-\varepsilon} = \sum_{i=1}^\infty \varepsilon^i,
\end{equation}
and note that the right side of Eq. \ref{Tau_A} can be rewritten as
\begin{equation} 
\beta \tau_A = \frac{1}{N} \sum_{i=1}^\infty N \varepsilon^i W_i 
=  W \sum_{i=1}^\infty i \varepsilon^i,
\label{tau_A1} 
\end{equation}
where $W_i$ has been replaced by $i W$ from Eq. \ref{T_i}. Therefore, 
\begin{equation}
\beta (\tau_A - W) = W \sum_{i=1}^\infty \left( i \varepsilon^i -  \varepsilon^i \right) 
			= W \sum_{i=1}^\infty (i-1) \varepsilon^i  \nonumber \\
			= W \sum_{j=0}^\infty j \varepsilon^{j+1} 
			=  \varepsilon W \sum_{j=1}^\infty j \varepsilon^j 
			= \varepsilon \beta \tau_A, 
\end{equation}
where Eq. \ref{tau_A1} was used for the last step. Therefore, 
\begin{equation}
\tau_A = W/(1-\varepsilon)
\end{equation}
as given in the main text. Clearly this derivation is meaningless if $\varepsilon = 0$ or $1$ because $\tau_A$ must be a finite number greater than zero to agree with experiment, so 
\begin{equation}
0< \varepsilon < 1,
\label{key1}
\end{equation}
as pointed out in the main text.  

Similarly, from Eq. \ref{f_i} the distribution from which the apparent lifetimes are drawn is given in terms of the transition rate $\lambda_B$ by the weighted average distribution
\begin{equation}  
f_A(\lambda_B, t) =  \frac{1}{N}\sum_{i=1}^\infty N_i f_i(\lambda_B, t)
=  \frac{1}{N \beta}\sum_{i=1}^\infty N \varepsilon^i f_i(\lambda_B, t)
= \frac{\lambda_B \varepsilon}{\beta} e^{-\lambda_B t} \sum_{i=1}^\infty \varepsilon^{i-1} 
	\frac{(\lambda_B t)^{i-1}}{(i-1)!}.           
\end{equation}
Therefore,
\begin{equation}  
f_A(\lambda_B, t) 
= (1-\varepsilon) \lambda_B e^{-\lambda_B t} \sum_{j=0}^\infty \frac{(\varepsilon \lambda_B t)^j}{j!}
= (1-\varepsilon) \lambda_B e^{-\lambda_B t} e^{\varepsilon \lambda_B t} 
= (1-\varepsilon) \lambda_B e^{-(1-\varepsilon) \lambda_B t}
= f_1(\lambda_A,t),        
\end{equation}
as given by Eq. \ref{D_A} in the main text, where $\lambda_A$ is defined as 
\begin{equation}
\lambda_A = (1-\varepsilon) \lambda_B.
\label{key2} 
\end{equation}

Finally, from Eq. \ref{S_i} the apparent survival function of each isolated particle in the sample is given in terms of the transition rate $\lambda_B$ by  
\begin{equation} 
S_A(\lambda_B t) = \frac{1}{N} \sum_{i=1}^\infty N (1- \varepsilon) \varepsilon^{i-1} S_i(\lambda_B, t) 
= (1-\varepsilon) e^{-\lambda_B t} \sum_{i=1}^\infty \varepsilon^{i-1} \sum_{n=1}^i \frac{(\lambda_B t)^{n-1}}{(n-1)!}.    
\end{equation} 
\begin{eqnarray}
S_A(\lambda_B t) & = & e^{-\lambda_B t} (1-\varepsilon) [1] + \nonumber \\
 &  & e^{-\lambda_B t} (1-\varepsilon) \varepsilon [1 + u] + \nonumber \\
 &  & e^{-\lambda_B t} (1-\varepsilon) \varepsilon^2 [1 + u + u^2/2] + \nonumber \\
 &  & e^{-\lambda_B t} (1-\varepsilon) \varepsilon^3[1 + u + u^2/2 + u^3/(3\cdot2)] + \nonumber \\
 &  & e^{-\lambda_B t} (1-\varepsilon) \varepsilon^4 [1 + u + u^2/2 + u^3/(3\cdot2) + u^4/(4\cdot3\cdot2) ] + \nonumber \\
 &  & \cdots,  
\end{eqnarray}
where, $u=\lambda_B t$.  Summing down the columns in the above equation gives
\begin{equation}
S_A(\lambda_B t) = e^{-\lambda_B t} (1-\varepsilon) \sum_{j=0}^\infty \varepsilon^j \left[ 1 + \varepsilon u  
+ \frac{(\varepsilon u)^2}{2}
+ \frac{(\varepsilon u)^3}{3\cdot2} 
+ \frac{(\varepsilon u)^4}{4\cdot3\cdot2}  
+ \cdots \right] 
	= e^{-\lambda_B t} \left[ \sum_{n=0}^\infty \frac{(\varepsilon u)^n}{n!} \right].  
\end{equation}
Replacing $u$ by $\lambda_B t$, and replacing the series by its sum gives
\begin{equation} 
S_A(\lambda_B t) = e^{-\lambda_B t} e^{\varepsilon \lambda_B t} 
		= e^{-(1-\varepsilon) \lambda_B t}
		= S_1((1-\varepsilon) \lambda_B t)
		= e^{-\lambda_A t}   
\end{equation}
as given by Eq. \ref{P_A} in the main text.  
 
\end{widetext}

\end{document}